# A Polarized Electron RF Photoinjector Using the Plane-Wave-Transformer (PWT) Design[*]

J. E. Clendenin, R. E. Kirby

*Stanford Linear Accelerator Center, 2575 Sand Hill Road, Menlo Park, CA 94025*

D. Yu, D. Newsham

*DULY Research Inc., 1912 MacArthur Street, Rancho Palos Verdes, CA 90275*

**Abstract**

Polarized electron beams are now in routine use in particle accelerators for nuclear and high energy physics experiments. These beams are presently produced by dc-biased photoelectron sources combined with rf chopping and bunching systems with inherently high transverse emittances. Low emittances can be produced with an rf gun, but the vacuum environment has until now been considered too harsh to support a negative electron affinity GaAs photocathode. We propose to significantly improve the vacuum conditions by adapting a PWT rf photoinjector to achieve reasonable cathode emission rates and lifetimes. This adaptation can also be combined with special optics that will result in a flat beam with a normalized rms emittance in the narrow dimension that may be as low as $10^{-8}$ m.

*Submitted to Physical Review Special Topics–Accelerators and Beams*

---

[*] Work supported in part by Department of Energy contract DE-AC03-76SF00515 (SLAC) and in part by DOE SBIR grant no. DE-FG03-96ER82156 (DULY Research Inc.).

# A Polarized Electron RF Photoinjector Using the Plane-Wave-Transformer (PWT) Design[*]

J. E. Clendenin, R. E. Kirby

*Stanford Linear Accelerator Center, 2575 Sand Hill Road, Menlo Park, CA 94025*

D. Yu, D. Newsham

*DULY Research Inc., 1912 MacArthur Street, Rancho Palos Verdes, CA 90275*

Polarized electron beams are now in routine use in particle accelerators for nuclear and high energy physics experiments. These beams are presently produced by dc-biased photoelectron sources combined with rf chopping and bunching systems with inherently high transverse emittances. Low emittances can be produced with an rf gun, but the vacuum environment has until now been considered too harsh to support a negative electron affinity GaAs photocathode. We propose to significantly improve the vacuum conditions by adapting a PWT rf photoinjector to achieve reasonable cathode emission rates and lifetimes. This adaptation can also be combined with special optics that will result in a flat beam with a normalized rms emittance in the narrow dimension that may be as low as $10^{-8}$ m.

## I. INTRODUCTION

Polarized electron beams are now a principal element of the physics program at a number of major accelerator centers including SLAC, JLAB, MIT/Bates, and Mainz. In addition, the SLC/SLD demonstrated that polarized electrons will be essential for future electron-positron colliders. In fact, using polarized electrons to measure the left-right asymmetry in electron-positron collisions at the $Z^0$ pole, the SLD has made the single most precise measurement of $\sin^2\theta_W$ to date despite recording significantly fewer events than any of the three competing LEP experiments [1]. Today, all polarized electron beams for accelerators are generated by dc-biased electron guns that utilize GaAs as the photocathode. The relatively long pulse—on the order of nanoseconds—generated by these sources is rf chopped and bunched in the injector to derive the desired pulse structure, including microbunch number and temporal width, to match the accelerator and experiment requirements.

The normalized rms transverse emittance of rf-bunched beams is typically on the order of $10^{-4}$ m. Future colliders such as NLC, JLC, TESLA, and CLIC, require an emittance of $\sim 10^{-8}$ m in at least one plane. The present collider designs achieve this extremely low emittance in the vertical plane using an appropriately designed damping ring. RF photocathode guns were introduced in 1985 by Fraser, Sheffield, and Gray [2]. Since the photoemitted electrons are rapidly accelerated to relativistic energies by electric fields that are much higher than used in dc guns, the effects of space charge on emittance growth are minimized. In 1989 Carlsten pointed out that since the initial emittance growth in an rf gun is correlated, this growth can be reversed by placing a solenoidal field immediately after the cathode [3]. An emittance-compensated rf photoinjector is normally designed to achieve the minimum emittance at a compensation point some distance beyond the solenoid exit. Simulations indicate that emittances as low as $10^{-6}$ m for 1 nC of charge per micropulse can be achieved with an rf photoinjector for round beams, although the measured values tend to be several times larger [4].

---

[*] Work supported in part by Department of Energy contract DE-AC03-76SF00515 (SLAC) and in part by DOE SBIR grant no. DE-FG03-96ER82156 (DULY Research Inc.).

Today, rf photoinjectors are in widespread use in FELs, in some low-energy linacs such as the Accelerator Test Facility (ATF) at BNL, and in a few larger accelerators such as the CLIC Test Facility (CTF) at CERN and the TESLA Test Facility (TTF) at DESY that are designed for R&D for future colliders.

Rf photoinjectors have been proposed as a source of cw unpolarized electron beams for energy recovery linacs (ERL) [5]. The gun laser required for an ERL may only be feasible if a GaAs (visible laser) or $CsK_2Sb$ (green) are utilized [6]. In this case the challenge for the PWT injector would be to provide adequate cooling. The cooling requirement is somewhat less stringent in some versions of electron ion colliders (EIC), which require polarized electrons, for which the rf frequency of the cw injector can be quite low [7].

The remainder of this paper is devoted to the possibility of producing polarized electron beams using an rf gun.

## II. MOTIVATION

While the emittance that is achieved directly with an rf photoinjector is not low enough to eliminate the damping ring for a future collider, it is significantly lower than that produced by rf bunchers. Thus with an rf gun the damping ring can be simpler and potentially less expensive. In addition, the whole rf bunching system is eliminated, although a laser system is gained.

Recently a technique for transforming a round beam having an angular momentum into a "flat" beam was proposed by Brinkmann, Derbenev, and Flöttmann [8] based on an optical transformation introduced earlier by Derbenev [9]. The necessary angular momentum for the beam can be provided by immersing the rf gun in a solenoidal magnetic field. The expected emittance in one plane is as low as several times $10^{-8}$ m. A proof-of-principle experiment to produce a flat beam in this manner has been successfully carried out at FNAL [10]. Thus it appears with this optical transformation technique, a beam could be produced using an rf photoinjector that would have an emittance low enough to eliminate completely the need for a damping ring for at least some collider designs and in particular for the TESLA design.

Consequently, since the electron beam for a collider must be polarized, it is now worth looking at the requirements for producing a polarized electron beam using an rf gun.

### III. REQUIREMENTS FOR A POLARIZED ELECTRON SOURCE

The requirements for producing a polarized electron beam using an rf gun have been discussed elsewhere [11] and are summarized below.

*Cathode.* Of the various III-V semiconductors that have been used for producing polarized electrons, GaAs is the most studied. The response time of bulk (typically 300—600 μm) GaAs photocathodes is known to approach nanoseconds, whereas for low emittance beams, the electron bunch in an rf gun should be launched with a bunch length that is <20° of rf phase at the cathode, i.e., for S-band <20 ps. The photon absorption depth for GaAs is on the order of a micron. Thus electrons promoted into the conduction band deep in the crystal will require a finite time to diffuse to the surface (during which time they thermalize to the bottom of the conduction band), resulting in bunch durations of nanoseconds even if a very short laser pulse is used. The bunch duration can be limited by using a thin cathode epilayer or by preventing diffusion electrons from escaping from the surface. Response times as low as a couple picoseconds have been observed for low-charge bunches generated by a sub-picosecond laser beam impinging on a thin epilayer (<200 nm) of GaAs [12].

*Activation.* GaAs photocathodes must be carefully activated to form a negative electron affinity (NEA) or nearly NEA surface. Activation requires the deposition of cesium on the perfectly clean GaAs surface such that when the quantum efficiency (QE) first peaks there will be about a half monolayer (ML, defined as one adatom per surface atom) of surface Cs. Immediately following the first QE peak, a mixture of Cs+oxide is applied until a second QE peak is achieved, whereupon the surface thickness of the Cs-oxide is about 1 ML. For a well-activated surface, the oxide generally ends up on the bottom of the added layer, while at the surface there is as much as ½ ML of free Cs. A zero electron affinity surface would block emission to vacuum of thermalized electrons that reach the surface from deep in the epilayer, thus shortening the bunch duration. Zero or even positive electron affinity surfaces might be desirable in some cases to reduce the initial bunch duration. With the possible exception of some types of field-assisted emission, methods of reducing the electron bunch duration result concomitantly in a reduced QE.

*Vacuum.* There is now a considerable amount of evidence that the time during which the QE of a GaAs photocathode decreases by a factor of 2 will be on the order of days or more only if the vacuum in which the cathode is operating is extremely good. It is clear that some gas species are more harmful than others. In general, any oxide with a partial pressure $>10^{-12}$ Torr will be a problem. Some gas species, such as $CO_2$, commonly found in UHV systems do not seem to affect the QE at all. The presence of other species, such as $O_2$ and $H_2O$, result in an increasingly rapid loss of QE when their partial pressures increase above $\sim 5 \times 10^{-13}$ Torr. The QE of a cathode aged in a UHV system can often be restored by adding more cesium to the surface except in the case of contamination by $H_2O$. These issues are summarized in reference [13]. Stainless-steel UHV chambers with residual pressures of $10^{-11}$ Torr, dominated by $H_2$ (which is relatively harmless), can readily be produced. There is much less experience with UHV all-copper systems.

Thin protective coatings have been found to protect some vacuum-sensitive photocathodes with very little loss of QE [14]. A similar coating might be found that would protect GaAs.

*High Voltage*. To extract the high charge needed for accelerator beams, the cathode must be biased at high voltage. Conventional guns use either dc or pulsed voltage. For these guns, the rate of decay of the cathode QE has been found to increase when the average dark current exceeds 100 nA. RF guns rely on the peak rf fields at the cathode surface to extract the charge. The rf fields produce dark currents, both internal and external to the rf cavity, that can readily destroy the cathode QE [15]. The external dark current (meaning the field-emitted electrons that exit the cathode rf cavity) in a conventional S-band rf gun can be as low as 1 nC per µs of rf [16]. It goes without saying that rf breakdowns—which produce extremely high currents, presumably after formation of a plasma source [17]—must be completely eliminated.

*Ion Bombardment*. Molecules from the residual gas in the rf chamber or desorbed by electron interception at the walls of the cavity can be ionized by the electron beam, then accelerated to the cathode causing serious deterioration to the cathode QE. This is a now well-known phenomenon for high duty-factor dc guns [18]. It is not clear that it will be as serious a problem in an rf gun, where the field direction is rapidly reversing.

*Magnetic Field*. Polarized photoelectrons extracted from a GaAs cathode have their polarization vector aligned parallel or anti-parallel to the axis of the excitation laser, which in this case is by design the momentum axis of the electron beam. Thus the beam is born with longitudinal polarization, and consequently axial magnetic fields have no effect on the polarization of the electron beam. Transverse magnetic fields will cause the electron-beam polarization vector to precess with respect to the momentum, which does not decrease the polarization if the energy spread is very low. In the latter case the polarization can be restored to longitudinal if desired. However, in an rf gun, during the emission of the electrons, there is a transverse rf magnetic field at the surface of the GaAs crystal where the relative energy spread of the electrons is high due to the rapidly increasing amplitude of the longitudinal rf electric field. Thus some depolarization is expected in an rf gun as the beam leaves the crystal surface. These rf fields also penetrate the crystal itself.

Except for very high-current beams, the cathode for an electron source is normally located in a magnetic-field free region to avoid the emittance dilution associated with a longitudinal magnetic field [19]. Embedding the cathode in a solenoidal field in order to produce a flat beam at the injector exit [8] should not affect the polarization vector of the extracted beam, but may have an effect inside the crystal. The spin direction of conduction-band electrons will precess around an external magnetic field as the electrons diffuse toward the crystal surface, i.e., introduce some mixing of the two adjacent spin states, which may result in a significant depolarization somewhat in analogy to the well-known D'yakonov-Perel' (DP) depolarization mechanism [20]. But since an rf gun requires a relatively short bunch length, most of the extracted electrons will be those that are initially promoted to the conduction band at a distance from the surface shorter than their mean free path (~1000 Å) from where they are rapidly accelerated, parallel to the solenoidal field, by the rf electric-field gradient that penetrates the crystal. (At 3 GHz, the skin depth for highly p-doped GaAs is ~50 µm [21].) Thus many of the conduction band electrons enter the band bending region (BBR, final 100 Å in this case) with a relatively high momentum. The additional accelerating field provided in the BBR can be up to 50 MV/m depending on the degree of electron affinity. However, even in the absence of external fields, the character of the electron kinetics in the BBR is complex and not well understood [22]. For an NEA surface with no external fields, it is known that there is a high probability that electrons will be trapped in the BBR where they lose some energy before emission to vacuum. One can expect some depolarization of trapped electrons in the presence of an external magnetic field. On the other hand, this depolarization effect is expected to be almost completely eliminated if the surface has a zero or positive electrons affinity, since the electrons emitted into vacuum are then almost entirely "hot" electrons, i.e., electrons that have undergone no collisions.

### IV. THE INTEGRATED PWT PHOTOINJECTOR

The plane wave transformer (PWT) accelerating structure, first proposed in the 1960s [23], is being successfully utilized in the Neptune Laboratory at UCLA as an S-band booster for a 1.6-cell rf gun [24]. More recently, a version of the PWT S-band structure that incorporates a photocathode has been designed and built by DULY Research Inc., and installed in the PEGASUS Laboratory at UCLA [25]. An X-band version is being studied [26].

The PWT design, shown schematically in Fig. 1, is a $\pi$-mode, standing-wave, linac structure, which differs from the conventional disk-loaded linac design in that the accelerating cells are formed by disks that are suspended and cooled by water-carrying rods in a large cylindrical tank. The plane-wave-transformer derives its name from the fact that rf power is coupled first into the annular region of the cylindrical tank outside the disk assembly in a TEM-like mode, and then coupled to the accelerating cells in a standing-wave, TM-like mode on axis. The cathode of the integrated photoinjector is located at one of the end plates and is an integral component of the photoelectron linac. A compact magnet is used to focus the electron beam after its creation from the photocathode, to preserve the beam emittances through the photoinjector linac, and to produce a moderate-energy electron beam of superior quality. Electrons are produced with a laser beam impinging on the photocathode at a small angle with respect to the beam axis.

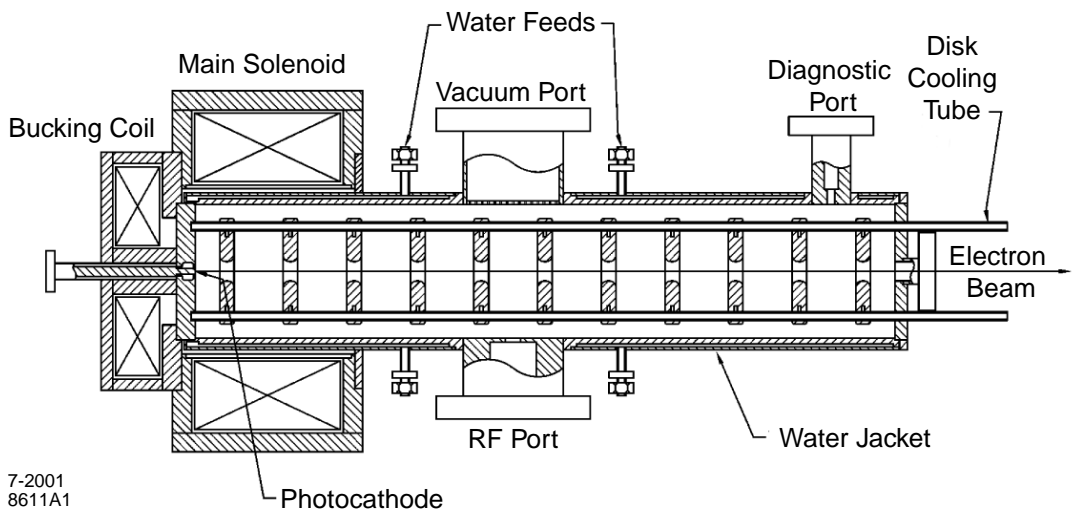

FIG. 1. Schematic of the DULY S-band integrated PWT photoelectron linac.

The design parameters of the PEGASUS photoinjector are shown in Table 1. The rf, beam dynamics and thermal/mechanical performance of the photoinjector have been simulated using 2D and 3D numerical codes (GdfidL, Parmela and COSMOS/M). The rf performance has been verified experimentally with full rf power (20 MW) with no breakdown. The design accelerating gradient of 33 MV/m of the 10+2/2 cell gun was based on an unloaded Q of 25000, while the measured Q was about half this value. Possible causes of the Q degradation are the multiple braze joints of the support rods and the copper coating of the stainless steel tank of the PWT. Beam tests are scheduled for 2001 as the PEGASUS laser system is nearing final completion. The photoinjector will be used to drive an undulator to produce infrared radiation.

TABLE I. S-Band PWT design parameters.

| Parameter | Value |
|---|---|
| Frequency | 2856 MHz |
| Energy | 20 MeV |
| Charge per Bunch | 1 nC |
| Normalized Emittance | 1 µm @ 60 MV/m |
| Energy Spread | <0.1% |
| Bunch Length (rms) | 2 ps |
| Rep Rate | 5 Hz |
| Peak Current | 100 A |
| Linac Length | 58 cm |
| Beam Radius | <1mm |
| Peak B Field | 1.8 kG |
| Peak Gradient | 60 MV/m |
| Peak Brightness | $2 \times 10^{14}$ A/(m-rad)$^2$ |

The interior construction of the PWT linac is shown in Fig. 2, while Fig. 3 shows the exterior of the assembled photoinjector.

### V. APPLICATION OF PWT DESIGN FOR A POLARIZED ELECTRON SOURCE

The PWT design for an integrated photoinjector offers additional advantages over the standard rf photoinjector for the purposes of producing a polarized electron beam. For a peak rf field at the cathode of 60 MV/m (Table 1), simulations indicate the normalized rms emittance for a 1-nC beam from the PWT S-band photoinjector should be about $1\times10^{-6}$ m [27], while at least 120 MV/m is required to achieve this same result with the split photoinjector design [28]. Positive fields on the order of ~30 MV/m and higher at the surface of a highly p-doped GaAs crystal will induce an inversion layer at the surface, which may allow field emission from the GaAs cathode itself. However, unlike the dc-field case, the formation of a strong inversion layer, which begins at ~60 MV/m, is limited by the inability of the minority carriers (in this case electrons) to keep up with the rf-field variations. Nonetheless, it may be necessary to limit even rf fields to below 60 MV/m.

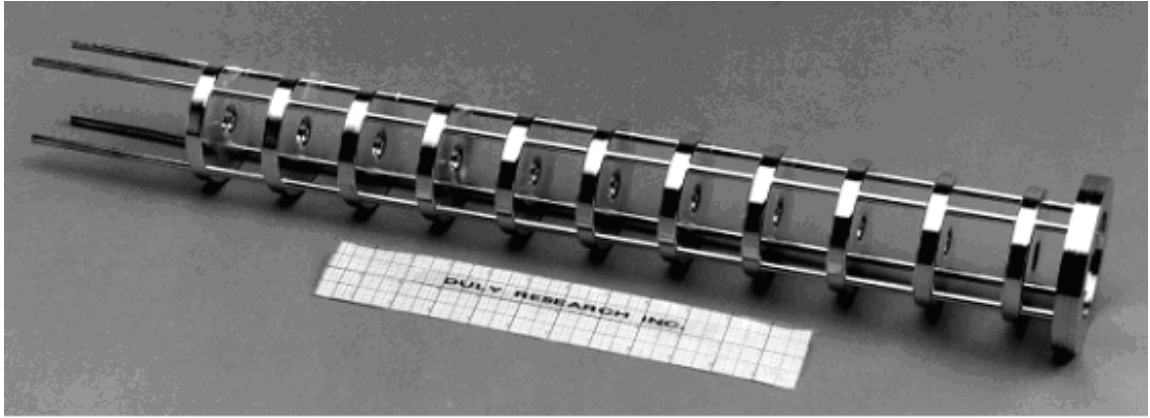

7-2001
8611A2

FIG. 2. Interior view of the S-band PWT accelerating structure.

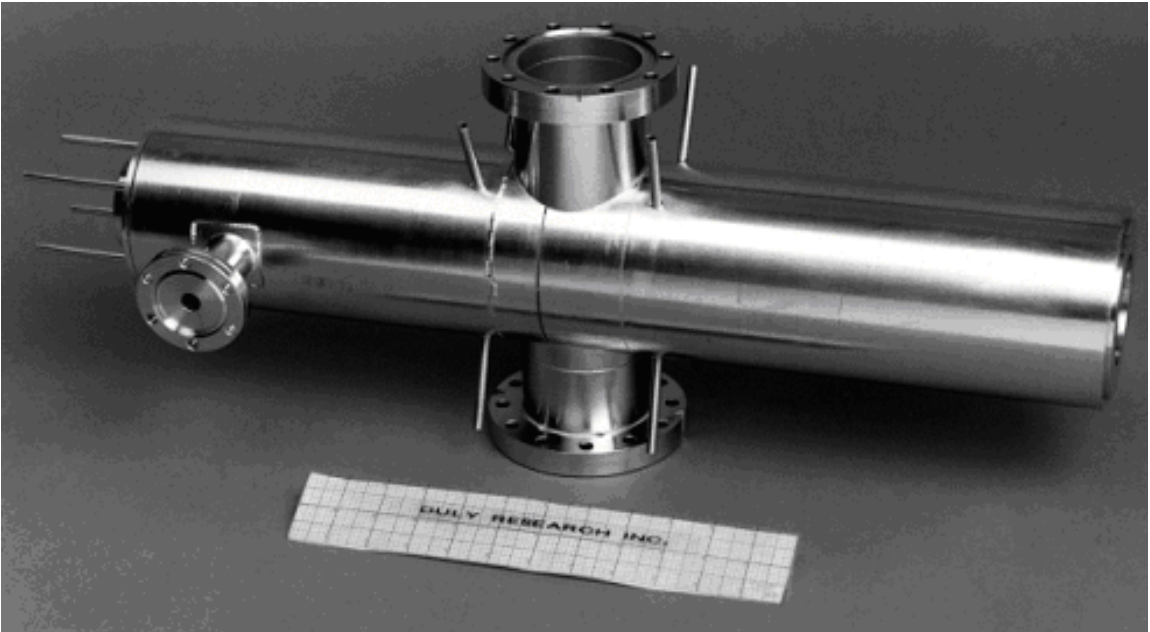

7-2001
8611A3

FIG. 3. Exterior view of the S-band PWT.

The separation of the tank wall from the accelerating structure in the PWT design allows for more and larger vacuum openings with high vacuum conductance that will help in the UHV requirement for the polarized electron source. Rather than requiring that the pumping occurs only through the iris of the disk, the PWT utilizes the large volume between the disks as the primary pumping pathway. In addition to increased pumping efficiencies, the large tank provides greater stored energy that can be used in a multi-pulse operation. This feature of the DULY photoinjector has resulted in a proposed application in CTF3 [29] and as the electron gun in an x-ray source based on Compton backscattering [30].

Recent discussion of voltage breakdown in accelerating structures [17] has led to the idea that the filling time of the accelerating structure can play an important role in the breakdown process. A longer filling time and a shorter structure, with a wider bandwidth (difference between the 0-mode and the operating mode), may mitigate the rf voltage breakdown process. These features are generally associated with standing wave structures. For the PWT structure, the separation between the accelerating structure and the tank wall allows for a wide range of cavity Q values, and hence, a wide range in filling times.

In the integrated PWT S-band linac, the tank wall and cooling rods were constructed of copper-plated stainless steel. As a part of this design, the decrease in the Q of the structure was studied as a function of relative degradation of the resistivity of the copper plating on these surfaces [31]. Fig. 4 shows the results of this study. A 50% decrease in the value of Q would be expected if the wall and rod resistivity increased by an order of magnitude.

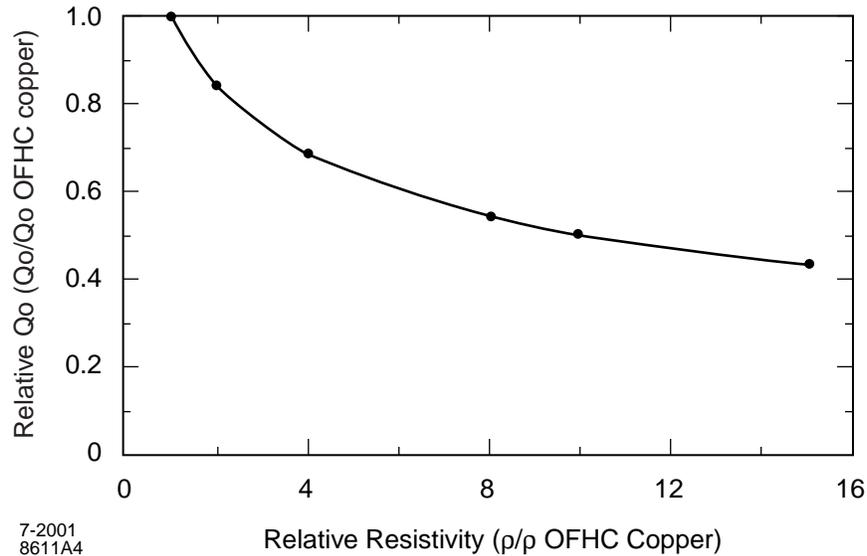

FIG. 4. Relative degradation of the $Q_o$ value in the DULY S-Band PWT as the resistivity of the tank wall and cooling rods is increased.

# VI. POSSIBLE MODIFICATIONS TO ENHANCE SUITABILITY OF PWT STRUCTURE AS A POLARIZED ELECTRON SOURCE

*Material and production quality.* The modifications to the existing PWT design that are required for use as a polarized electron source can be separated into two categories: design modifications, and material and production-quality improvements. Recent tests indicate that to minimize dark current and rf breakdown, Class 1 OFHC copper forged using the hot isostatic pressure (HIP) method and single-point diamond machining to a surface roughness of 0.05 μm or better should be used for fabrication, followed by simple rinsing in ultra-pure water [32]. Maintenance of the ultrahigh vacuum ($< 10^{-10}$ Torr) would probably necessitate assembly in a Class 100 or better clean room.

*Vacuum pumping.* To better improve the vacuum, the tank wall can be coated with a thin (couple micrometers) film of TiZr (or TiZrV), a recently proposed getter material that can be reactivated at relative low temperatures several times after exposure to atmosphere [33], which should be adequate for a load-locked system (see below). External getter (NEG) pumps have been used successfully to reduce the vacuum in an operating 1.6-cell S-band rf gun by an order of magnitude [34] and should be used here as well. Recently the GaAs photocathode of a dc-biased gun was extended into a chamber containing a massive array of NEG pump elements [18]. The resulting dramatic increase in pumping speed at the cathode significantly improved the gun performance. The thin TiZr films are almost transparent to rf [35]. Consequently, as indicated in Fig. 4, the effect of the film on the cavity Q is expected to be very small. The maximum electric field at the tank wall is about 56% of the maximum on axis and located at the center line of a disk. Field emission related to the coating can be minimized by carefully applying a very smooth layer.

*Cooling.* Simulations indicate that for the present DULY S-band PWT design, a klystron power of 25 MW at 10 Hz and with a flow rate of 1 liter/min in the disk, the thermal gradient across the disk is about 0.25°C, and the cold point on the disk is 1°C warmer than the cooling water. If the repetition rate is increased to 180 Hz, which increases the average power by a factor of 18, it will be necessary to increase the flow rate through the disk by a factor of 2 (4) to reduce the temperature difference between disk and water to 6 (2) °C while the gradient across the disk will be 4.3°C independent of flow rate. Additional cooling is not expected to be needed due to the added getter material alone.

*PWT dimensions.* The flexibility of the dimensions of the PWT design permits modifications that will enhance multi-bunch acceleration and vacuum pumping. The former may necessitate a shorter overall length for a given rf power and an enlarged beam pipe to minimize beam breakup (BBU) and a larger rf power coupling for heavy beam loading. Increasing the tank radius will not only increase the stored energy and the Q, but will also increase the conductance, thus allowing more efficient vacuum pumping. In addition, the first half-cell can be enlarged to a 0.625-cell, as is now done with separated-function guns, to decrease the emittance.

*Load-lock.* Operation of the integrated PWT photoinjector as a polarized electron source requires the use of a GaAs cathode. The current DULY design uses a de-mountable copper cathode. This cathode system was designed for months of stable operation between cathode replacements; however, it requires that the entire photoinjector be brought to atmosphere for the replacement, resulting in a long operational down-time because of the need to re-evacuate and bake the system after the cathode replacement. Regular replacement of the GaAs cathode will require a load-lock for inserting and removing cathodes without breaking the gun vacuum, since GaAs, once activated, must be maintained always in an ultra-high vacuum.

The proposed load-lock is shown in Fig. 5 with a cathode plug (1, the figure) exchange in progress. The entire system consists of a process chamber (not shown) where cathode heat cleaning and cesiation are executed and where new cathodes are first loaded for activation prior to exchange in the main load-lock. The main load-lock is computer-controlled for remote withdrawal, for touch-up cesiation (14), and for re-insertion of a cathode into the photoinjector. The cathode transport device (2,5) is a rack-and-pinion gear mechanism. The advantage of its use, over bellows-sealed or magnetically-coupled motions, is the ability to sense, via appropriate mechanical pressure gauging on the pinion rotary motion the load applied to the cathode when the cathode is seated in the injector (13). This capability ensures that the rack (2) exerts a constant force on the cathode, thereby avoiding an operational tune shift.

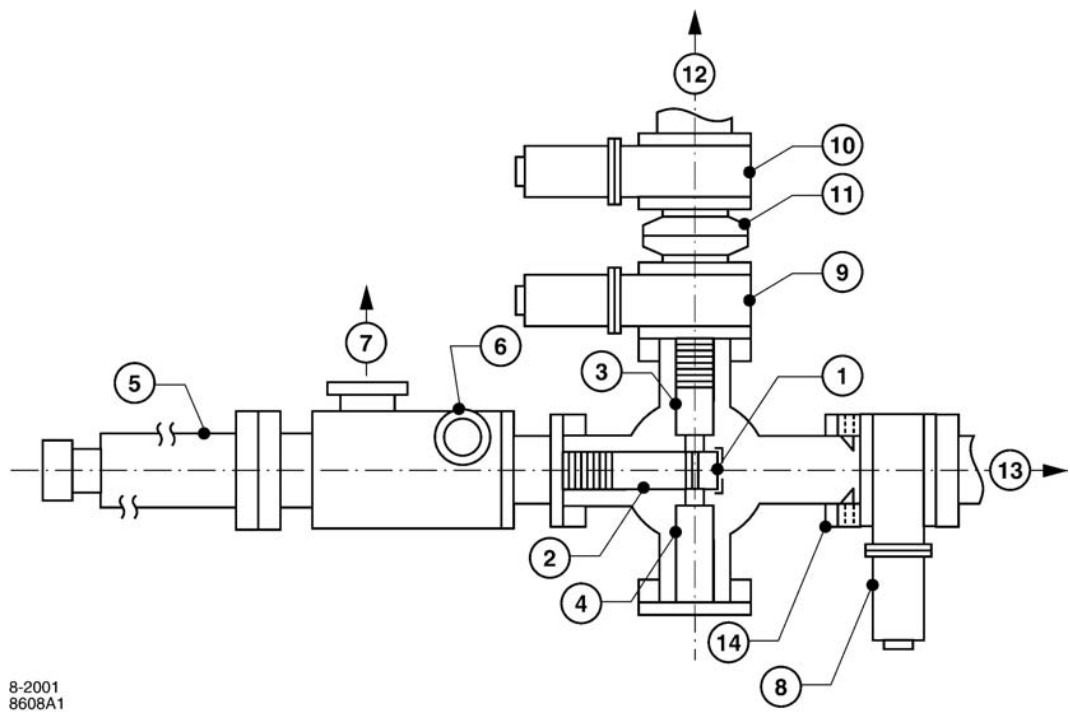

FIG. 5. Load-lock system, depicting cathode plug exchange in progress: 1) photocathode plug, 2) gun cathode rack, 3) cathode exchange rack, 4) cathode exchange support anvil, 5) rack and pinion gun loader, 6) motorized drive, 7) to pumping system, 8) gun isolation valve, 9) load-lock isolation valve, 10) process/exchange unit isolation valve, 11) quick-disconnect flange, 12) to process/activation stations, 13) to photoinjector, 14) cesiation unit (channel cesiators, laser window and laser, photocurrent detector).

When a cesiation touch-up is required, the cathode is retracted through the isolation valve (8) to a position adjacent to the cesiation unit (14). The isolation valve is closed remotely for safety and a cesiation cycle is performed with a local diode laser and current collector. The valve is then re-opened and the cathode re-injected into the gun. A new cathode requires the complete removal of the old through a quickly attached flange (11) using a cathode exchange unit (12).

## VII. CONCLUSIONS

RF guns can produce low emittance beams for particle accelerators. Present and future accelerators require polarized electron beams, but to date polarized electron beams for accelerators have been produced only with dc-biased or pulsed guns utilizing GaAs photocathodes. Maintaining emission from a GaAs cathode requires an extremely good vacuum, which is not available with a standard rf gun. We have shown the PWT design to be uniquely adaptable to the production of low-emittance polarized electron beams. The salient features are better vacuum and lower rf fields for optimum emittance. The former is a product of the open structure and versatility of dimensions in the PWT design. The latter is primarily a feature of the integrated or multi-cell injector in which there is no drift space between the cathode and first associated cells (gun) and the initial accelerating cells (booster).